\documentclass[reprint,amsmath,amssymb,aps,]{revtex4-2}

\usepackage{graphicx}	
\usepackage{dcolumn}	
\usepackage{bm}			
\usepackage{hyperref}
\hypersetup{
    colorlinks,
    citecolor=blue,
    filecolor=black,
    linkcolor=blue,
    urlcolor=black
}
\begin{document}

\title{Composite Flavon-Higgs Models}

\author{Yi Chung}
\email{yichung@ucdavis.edu}
\affiliation{
 Center for Quantum Mathematics and Physics (QMAP), Department of Physics, \\University of California,  Davis, CA 95616, U.S.A.
}

\begin{abstract}
We consider a composite Higgs model based on the $SU(6)/Sp(6)$ coset, where an $U(1)$ subgroup of $Sp(6)$ is identified as the flavor symmetry. A complex scalar field $s$, which is a pseudo-Nambu-Goldstone boson of the broken symmetry, carries a flavor charge and plays the role of a flavon field. The $U(1)_F$ flavor symmetry is then broken by a VEV of the flavon field, which leads to a small parameter and generates the mass hierarchy between the top and bottom quarks. A light flavon below the TeV scale can be naturally introduced, which provides a fully testable model for the origin of flavor hierarchy. A light flavon also leads to substantial flavor changing neutral currents, which are strongly constrained by the flavor experiments. The direct search of additional scalar bosons can also be conducted in HL-LHC and future hadron colliders.
\end{abstract}

\maketitle

\section{Introduction}

The Standard Model (SM) of particle physics successfully describes all known elementary particles and their interactions. However, there are still a few puzzles that have yet to be understood, including two mysterious hierarchies. One is the well-known \textbf{hierarchy problem}. With the discovery of light Higgs bosons in 2012 \cite{Chatrchyan:2012xdj, Aad:2012tfa}, the last missing piece of the SM seemed to be filled, but SM does not address the UV-sensitive nature of scalar bosons. The Higgs mass-squared receives quadratically divergent radiative corrections from the interactions with SM fields, which require an extremely sensitive cancellation to have a $125$ GeV Higgs boson. The other puzzle is related to the large hierarchies in the masses and mixings of the SM fermions. Even within the quark sector, the masses of quarks span over six orders of magnitude. The mixing angles also show a hierarchical structure. The problem is known as the \textbf{flavor puzzle} \cite{Babu:2009fd}, which represents the mysterious structure of SM Yukawa couplings.

One such appealing solution to the hierarchy problem is the composite Higgs model (CHM), where the Higgs doublet is the pseudo-Nambu-Goldstone bosons (pNGB) of a spontaneously broken global symmetry of the underlying strong dynamics~\cite{Kaplan:1983fs, Kaplan:1983sm}. Through the analogy of the chiral symmetry breaking in quantum chromodynamics (QCD), which naturally introduces light scalar fields, i.e., pions, we can construct models with light Higgs bosons. In a CHM, an approximate global symmetry $G$ is spontaneously broken by some strong dynamics down to a subgroup $H$ with a symmetry breaking scale $f$. The heavy resonances of the strong dynamics are expected to be around the compositeness scale $\sim 4\pi f$. The pNGBs of the symmetry breaking, on the other hand, can naturally be light with masses $<f$ as they are protected by the shift symmetry.

For the flavor puzzle, the hierarchy in the masses and mixings of the SM fermions can be achieved by assuming an abelian $U(1)_{F}$ flavor symmetry \cite{Froggatt:1978nt}, where different SM fermions carry different charges. The low-energy effective Yukawa coupling terms require the insertion of additional scalar fields as
\begin{equation}
\mathcal{L}_{\text{Yukawa}}=
y_{ij}\left(\frac{s}{\Lambda_F}\right)^{a_{ij}}\bar{q}_{L,i}{H}q_{R,j},
\end{equation}
where $y_{ij}$ is a $\mathcal{O}(1)$ coupling, the complex scalar field $s$ is called flavon field, and ${\Lambda_F}$ is the scale of flavor dynamics. After the flavon field acquires a VEV, it will lead to a small parameter $\epsilon=\langle s\rangle/\Lambda_F$ and result in the hierarchy of SM Yukawa couplings. It is known as the Froggatt-Nielsen (FN) mechanism. Despite the success in explaining the flavor structure, the scale of flavor dynamics is not predicted and can be arbitrarily high. Also, the flavon as a scalar boson receives large radiative corrections from the interactions with SM fields and is expected to be well beyond the collider search.

In this paper, we explore models that can address these two problems at once and provide predictive experimental signatures which can be probed by colliders. We choose the specific CHMs with the unbroken subgroup large enough to include the $U(1)_{F}$ symmetry. That is, the flavor symmetry arises as part of the accidental global symmetry of the strong dynamics. Under this construction, the Higgs doublet and the flavon are pNGBs of the spontaneously broken global symmetries. In this case, the hierarchy problem is relieved, and a light flavon is naturally introduced, which provides a testable theory for the origin of flavor hierarchy.

Efforts to generate flavons as pNGBs have been implemented in the little flavon model \cite{Bazzocchi:2003ug, Bazzocchi:2003vh}, which is aimed at realizing collective symmetry breaking on the flavon field. Versions combined with Higgs doublet were also studied \cite{Bazzocchi:2004qx, Bazzocchi:2004cv}, but the large symmetry group makes them uncompelling. They also failed to treat the generation of Yukawa coupling carefully. Other attempts aiming at generating the Higgs and flavon from a common source have been studied recently \cite{Alanne:2018fns}, inspired by axiflavon models \cite{Calibbi:2016hwq, Ema:2016ops}. However, the scalar flavon in the model is not the pNGB mode but the heavy unstable radial mode, which is hard to be detected, and the FN fields are elementary vector-like fermions added by hands. There are also other efforts to relate the flavor breaking scale to the electroweak scale but within the framework of 2HDM \cite{Bauer:2015fxa, Bauer:2015kzy}.

For a concrete model, we consider a composite Higgs model based on the $SU(6)/Sp(6)$ coset, where the unbroken $Sp(6)$ is large enough to include both the SM gauge group and the global flavor symmetry group $SU(2)_W\times U(1)_Y\times U(1)_{F}$. The flavons as well as two Higgs doublets are the pNGBs of the coset. We then show how a suppressed Yukawa coupling can be generated through partial compositeness with specific flavor charge assignments. We discuss different scenarios to realize the Froggatt-Nielsen mechanism and generate the top-bottom mass hierarchy. The experimental constraints of different cases will also be discussed.

\section{The $SU(6)/Sp(6)$ CHM}\label{sec:Model}

The $SU(6)/Sp(6)$ coset is one of the earliest cosets employed in little Higgs models \cite{Low:2002ws} where the collective symmetry breaking for the quartic term was realized. Recently, it was considered for dark matter study \cite{Cai:2018tet} and natural Higgs potential \cite{Cheng:2020dum}. It was pointed out in \cite{Cheng:2020dum} that there is a $U(1)$ Peccei-Quinn like subgroup \cite{Peccei:1977ur}, which protects the theory from dangerous tadpole terms and flavor changing neutral currents. In this paper, this subgroup is identified as $U(1)_F$ flavor symmetry to realize the Froggatt-Nielsen mechanism. For our purpose, we will focus on the fermion sector and Yukawa couplings in the main text. The gauge sector and the pNGB potential are discussed in the appendix. A more comprehensive discussion on these topics can also be found in \cite{Cheng:2020dum}.

\subsection{Basics of $SU(6)/Sp(6)$}

The $SU(6)/Sp(6)$ non-linear sigma model can be parametrized by a sigma field $\Sigma^{ij}$, which transforms as an anti-symmetric tensor representation $\mathbf{15}$ of $SU(6)$, where $i, j=1, \ldots 6$ are $SU(6)$ indices. The transformation under $SU(6)$ can be expressed as $\Sigma \to g\,\Sigma\, g^T$ with $g\in SU(6)$ or as $\Sigma^{ij} \to {g^i}_k{g^j}_\ell\Sigma^{k\ell}$ with indices explicitly written out. The scalar field $\Sigma$ has an anti-symmetric VEV $\langle \Sigma\rangle=  \Sigma_0^{\alpha\beta}$ (with $\alpha$, $\beta$ representing $Sp(6)$ indices), where
\begin{equation}
\Sigma_0=
\begin{pmatrix}
0  &  -\mathbb{I}_{3\times 3} \\
\mathbb{I}_{3\times 3}   &  0 \\
\end{pmatrix}.
\end{equation}
The $\Sigma$ VEV breaks $SU(6)$ down to $Sp(6)$, producing 14 Nambu-Goldstone bosons.

The 35 $SU(6)$ generators can be divided into unbroken ones and broken ones with each type satisfying
\begin{equation}
\begin{cases}
\text{unbroken generators}   &~T_a   : T_a\Sigma_0+\Sigma_0T_a^T=0~,\\
\text{broken generators}       &X_a   : X_a\Sigma_0-\Sigma_0X_a^T=0~.
\end{cases}
\end{equation}
The Nambu-Goldstone fields can be written as a matrix with the broken generators:
\begin{equation}
\xi(x)={\xi^i}_\alpha(x)\equiv e^{\frac{i\pi_a(x)X_a}{2f}}. 
\end{equation}
Under $SU(6)$, the $\xi$ field transforms as $\xi \to g \,\xi\, h^{\dagger}$ where $g \in SU(6)$ and $h \in Sp(6)$, so $\xi$ carries one $SU(6)$ index and one $Sp(6)$ index. 
The relation between $\xi$ and $\Sigma$ field is given by
\begin{equation}
\Sigma(x)=\Sigma^{ij}(x)\equiv \xi \,\Sigma_0\,\xi^T=e^{\frac{i\pi_a(x)X_a}{f}}\Sigma_0~.
\end{equation}
The complex conjugation raises or lowers the indices. The fundamental representation of $Sp(6)$ is (pseudo-)real and the $Sp(6)$ index can be raised or lowered by $\Sigma_0^{\alpha\beta}$ or $\Sigma_{0,\alpha\beta}$.

The broken generators and the corresponding fields in the matrix can be organized as follows ($\epsilon=i\sigma^2$):
\begin{equation}
\pi_aX_a=
\begin{pmatrix}
 \frac{\phi_a}{\sqrt{2}}\sigma^a-\frac{\eta}{\sqrt{6}}\bf{1}  &  ~H_2    &  ~\epsilon s     & ~H_1    \\
H_2^\dagger   &  ~\frac{2\eta}{\sqrt{6}}   &   ~-H_1^T  &    ~0 \\  
\epsilon^T s^*  &  ~-H_1^*    &    ~\frac{\phi_a}{\sqrt{2}}\sigma^{a*}-\frac{\eta}{\sqrt{6}}\bf{1}   &  ~H_2^*\\
H_1^\dagger   &  ~0   &   ~H_2^T  &   ~\frac{2\eta}{\sqrt{6}} \\  
\end{pmatrix}.
\end{equation}
In this matrix, there are 14 independent fields. They are (under $SU(2)_W$): a real triplet $\phi_a$, a real singlet $\eta$, a complex singlet $s$ (as the flavon field), and two Higgs (complex) doublets $H_1$ and $H_2$. We effectively end up with a two-Higgs-doublet model (2HDM). The observed Higgs boson will correspond to a mixture of $h_1$ and $h_2$ inside two Higgs doublets $H_1=H_{1/2}\supset \frac{1}{\sqrt{2}}\bigl(\begin{smallmatrix}0\\h_1\end{smallmatrix}\bigr)$ and $H_2=H_{-1/2}\supset\frac{1}{\sqrt{2}}\bigl(\begin{smallmatrix}h_2\\0\end{smallmatrix}\bigr)$. Using the $\xi$ and $\Sigma$ matrices, we can construct the low energy effective Lagrangian for the flavon field, the Higgs fields, and all the other pNGBs.

\subsection{Unbroken subgroups of $Sp(6)$}

To realize the FN mechanism, we need a global symmetry with scalars and fermions charged under it. Within the $Sp(6)$ symmetry, there are several unbroken $U(1)$ symmetries. The symmetries with generators
\begin{align}
\frac{1}{2}&
\begin{pmatrix}
\sigma^a   &  0  &  0  &  0 \\
0   &  0  &  0  &  0\\
0   &  0  &  -\sigma^{a*} &  0\\
0   &  0  &  0  &  0  \\  
\end{pmatrix}\text{ and }
\frac{1}{2}
\begin{pmatrix}
0_{2\times2}   &  0  &  0  &  0 \\
0   &  1  &  0  &  0 \\
0   &  0  &  0_{2\times2}  &  0 \\
0   &  0  &  0  &  -1  \\  
\end{pmatrix}+ X \mathbf{I}\nonumber
\end{align}
are identified as the SM gauge group $SU(2)_W$ and $U(1)_Y$, which are discussed in appendix~\ref{Gauge}.

Besides the SM gauge group, there is one more $U(1)_F$ global symmetry with the generator
\begin{equation}
U(1)_{F}: \frac{1}{2}
\begin{pmatrix}
\mathbb{I}_{2\times2}   &  0  &  0  &  0 \\
0   &  0  &  0  &  0 \\
0   &  0  &  -\mathbb{I}_{2\times2}   &  0 \\
0   &  0  &  0  &  0  \\  
\end{pmatrix}.\nonumber
\end{equation}
Under $U(1)_{F}$, the complex scalar field $s$ has charge 1, both Higgs doublets $H$ have charge 1/2, and other pNGB fields have charge 0. The complex singlet $s$ can then be identified as the composite flavon field. We then get the charge assignment for all pNGBs as 
\begin{equation}
s: 1, \quad H_1,~H_2: 1/2, \quad \phi,~\eta: 0~,
\end{equation}
which is a little different from the normal FN mechanism since Higgs also carries flavor charges
\footnote{In fact, this global symmetry is more similar to the $U(1)$ Peccei-Quinn (PQ) symmetry \cite{Peccei:1977ur}. Models that identify $U(1)_{PQ}$ as flavor symmetry had been studied in axiflavon models \cite{Calibbi:2016hwq, Ema:2016ops}. However, in this paper, we will not deal with the strong CP problem and axions, so we would like to call it $U(1)_{F}$ flavor symmetry.}.
So far, we get the desired scalar sector with the flavon and Higgs doublets. We can then move on to the fermion sector.

\section{Yukawa coupling}\label{sec:Yukawa}

In CHMs, the SM Yukawa couplings can arise from the partial compositeness mechanism~\cite{Kaplan:1991dc}. That is, elementary fermions mix with composite operators of the same SM quantum numbers from the strong dynamics,
\begin{equation}
\mathcal{L}=\lambda_{L}\bar{q}_{L}{O}_{R}+\lambda_{R}\bar{q}_{R}{O}_{L},
\end{equation}
where $q_L$, $q_R$ are elementary fermions and ${O}_L$, ${O}_R$ are composite operators of some representations of $SU(6)$.

To be able to mix with the elementary fermions, the representations of the composite operators must contain states with the same SM quantum numbers as the SM fermions. To account for the correct hypercharge, e.g., $q_L=2_{1/6}$ for left-handed quarks, $q_R=1_{2/3}$ for right-handed up-type quarks, and $q_R=1_{-1/3}$ for right-handed down-type quarks, the composite operators need to carry additional charges under the $U(1)_X$ outside $SU(6)$, and the SM hypercharge is a linear combination of the $SU(6)$ generator $\text{Diag}(0,0,1/2, 0,0, -1/2)$ and $X$.

Let us start with the top quark. To get the top Yukawa coupling, the suitable and economical choice of composite operators is $\mathbf{6}$ with $X=1/6$. The composite operator as a $\mathbf{6}_{1/6}$ of $SU(6)$ (where the subscript $1/6$ denotes its $U(1)_X$ charge) can be decomposed under the SM gauge group as
\begin{equation}
O_{L,R}^i\sim{\xi^i}_\alpha Q_{L,R}^\alpha\sim
\mathbf{6}_{1/6}=\mathbf{2}_{1/6}\oplus\mathbf{1}_{2/3}\oplus\mathbf{\bar{2}}_{1/6}\oplus\mathbf{1}_{-1/3},
\end{equation}
where $Q_{L,R}$ are the corresponding composite resonances. The composite states $Q_{L,R}$ belong to the $\mathbf{6}$ representations of $Sp(6)$ and play the roles of SM fermion composite partners. For $SU(2)$, $\mathbf{2}$ and $\mathbf{\bar{2}}$ are equivalent and related by the $\epsilon$ tensor. We make the distinction to keep track of the order of the fermions in a doublet. We see that the composite states have the appropriate quantum numbers to mix with the SM quarks.

The left-handed top quark can mix with the first two components of the sextet. The mixing term can be express as
\begin{equation}
\lambda_{L}\bar{q}_{La}{\Lambda^a}_iO_R^i=
\lambda_{L}\bar{q}_{La}{\Lambda^a}_i\left({\xi^i}_\alpha Q_{R}^\alpha \right),
\end{equation}
where $a$ represents an $SU(2)_W$ index, and 
\begin{equation}
{(\Lambda)^a}_i=\Lambda=
\begin{pmatrix}
1   &  0  &  0  & 0  & 0  & 0  \\
0   &  1  &  0  & 0  & 0  & 0  \\
\end{pmatrix}
\end{equation}
is the spurion which keeps track of the symmetry breaking.

To get the complete top Yukawa coupling, we couple the elementary right-handed top quark to the $\mathbf{\bar{6}}_{1/6}$, which decomposes under $SU(2)_W \times U(1)_Y$ as
\begin{equation}
O'_{L,Rj}\sim{{\xi^*}_j}^\beta \Sigma_{0\beta\alpha}Q_{L,R}^\alpha\sim
\mathbf{\bar{6}}_{1/6}=\mathbf{\bar{2}}_{1/6}\oplus\mathbf{1}_{-1/3}\oplus\mathbf{2}_{1/6}\oplus\mathbf{1}_{2/3}~.
\end{equation}
The right-handed top quark mixes with the last component of the $\mathbf{\bar{6}}_{1/6}$, which can be written as
\begin{equation}
\lambda_{t_R}\bar{t}_{R}{\Gamma_{t_R}}^jO'_{Lj}=
\lambda_{t_R}\bar{t}_{R}{\Gamma_{t_R}}^j\left({{\xi^*}_j}^\beta \Sigma_{0\beta\alpha}Q_{L}^\alpha \right),
\label{eq:right_partial}
\end{equation}
where $\Gamma_{t_R}=\left(0~0~0~0~0~1 \right)$ is the corresponding spurion.

Combining $\lambda_{L}$ and $\lambda_{t_R}$ couplings, we can generate the SM Yukawa coupling for the top quark
\begin{align}
&\sim 
\lambda_{L}\lambda_{t_R}\bar{q}_{La}{\Lambda^a}_i{\xi^i}_\alpha \Sigma_{0}^{\alpha\beta}{{\xi^T}_\beta}^j{\Gamma}^\dagger_{t_Rj} t_R\supset \lambda_{L}\lambda_{t_R}\left(\bar{q}_LH_2t_R\right)~.
\label{eq:top_yukawa}
\end{align}
The top quark gets its mass from the vacuum of $H_2$ as
\begin{align}
m_t=\frac{\lambda_{L}\lambda_{t_R}}{g_T}\frac{v_2}{\sqrt{2}}~,
\end{align}
where $g_T$ is a coupling of the composite top partners.

Similarly, for the bottom quark, we can couple $b_R$ to the third component of $\mathbf{\bar{6}}_{1/6}$ with the coupling $\lambda_{b_R}$ and spurion $\Gamma_{b_R}=\left(0~0~1~0~0~0 \right)$. This generates a bottom Yukawa coupling
\begin{align}
&\sim 
\lambda_{L}\lambda_{b_R}\bar{q}_{La}{\Lambda^a}_i{\xi^i}_\alpha \Sigma_{0}^{\alpha\beta}{{\xi^T}_\beta}^j{\Gamma}^\dagger_{b_Rj} b_R\supset \lambda_{L}\lambda_{b_R}\left(\bar{q}_LH_1b_R\right)~,
\label{eq:bottom_yukawa_0}
\end{align}
where the bottom quark gets its mass from the vacuum of $H_1$ instead.

In this paper, we will not address the lepton sector, so there are only two types of 2HDMs satisfying the natural flavor conservation \cite{Glashow:1976nt, Paschos:1976ay}. They are categorized by Type-I and Type-II based on the Yukawa couplings of the quarks. So far, the Yukawa couplings of the third generation quarks come from different Higgs doublets, which implies a \textbf{Type-II 2HDM}. The smallness of the bottom quark mass can be achieved by a small VEV of $H_1$, i.e. a large tan$\beta$ Type-II 2HDM. However, the parameter space with a large tan$\beta$ is strongly constrained by direct searches, and it is also not what we want. To get mass hierarchy between the top and bottom through the FN mechanism, we want an insertion of the flavon field $s$ in these Yukawa coupling terms.

\section{Froggatt-Nielsen mechanism}\label{sec:FNmech}

\subsection{FN mechanism: The first taste}

Before we move on to the correct FN mechanism setup, let us first look at the flavor charges of quarks. In the previous section, all the quarks are embedded in $\mathbf{{6}}_{1/6}$ and $\mathbf{\bar{6}}_{1/6}$ of $SU(6)$ without additional flavor charges, which are decomposed under $SU(2)_W \times U(1)_{F}$ as
\begin{align}
\mathbf{{6}}_{0}&=\mathbf{{2}}_{1/2}\oplus\mathbf{1}_{0}\oplus\mathbf{\bar{2}}_{-1/2}\oplus\mathbf{1}_{0}~,\\
\mathbf{\bar{6}}_{0}&=\mathbf{\bar{2}}_{-1/2}\oplus\mathbf{1}_{0}\oplus\mathbf{2}_{1/2}\oplus\mathbf{1}_{0}~.
\end{align}
It means that the flavor charges of fermions are set as 
\begin{equation}
q_L=(t_L,~b_L)^T: 1/2, \quad t_R,~b_R: 0~,
\label{eq:charge_assignment_1}
\end{equation}
where both right-handed quarks have no flavor charge.

Within this assignment, we can already write down a suppressed bottom quark mass through the FN mechanism by the term like
\begin{align}
\frac{1}{f}\left(\bar{q}_Ls\tilde{H}_2b_R\right)\sim
\left(\frac{v_sv_2}{2f}\right)\bar{b}_Lb_R~,
\label{eq:bottom_yukawa_01}
\end{align}
where $v_s$ is the VEV of the flavon field. The term satisfies the flavor symmetry. The reason it is possible is that the top quark gets mass from ${H}$ with flavor charge 1/2, but the bottom quark can get mass from $\tilde{H}$ with flavor charge $-1/2$. However, it turns out that this term can not successfully realize the FN mechanism in this model.

To see that, we can go back to the term we derived for the bottom quark mass in Eq.~\eqref{eq:bottom_yukawa_0}. In the non-linear Sigma model, if we expand the $\Sigma$ field to the next order, it becomes
\begin{align}
\label{eq:bottom_yukawa_02}
\bar{q}_L(H_1+\frac{i}{2f}s\tilde{H_2})b_R
\supset \frac{i}{2f}\left(\bar{q}_Ls\tilde{H}_2b_R\right)~,
\end{align}
which already contains the term in Eq.~\eqref{eq:bottom_yukawa_01}. That means, due to the shift symmetry of pNGBs, the term $s\tilde{H}_2$ can only show up following $H_1$. That also means we can always transfer the nontrivial vacuum of $\langle s\tilde{H}_2\rangle$ to the leading order $\langle H_1\rangle$ by shift symmetry. Therefore, the bottom quark mass still comes from $\langle H_1\rangle$, and it is equivalent to the Type-II 2HDM we have already gotten.

If we define $\Delta_f \equiv [f_L]-[f_R]$ as the difference between flavor charges of left-handed and right-handed fermions. Fixing the top quark charge as in Eq.~\eqref{eq:charge_assignment_1} with $\Delta_t=1/2$, we find that $\Delta_b=1/2$ gives us the bottom quark mass through $H_1$, which leads to a Type-II 2HDM. $\Delta_b=-1/2$, instead, generates the bottom quark mass through $\tilde{H}_2$ and makes it a Type-I 2HDM. Either case is just normal 2HDM. To realize the FN mechanism, we need to have a larger $|\Delta_b|$, which would allow us to generate the bottom Yukawa coupling term with the insertion of two pNGBs, $s$ and $H$, at the same time. That also requires us to embed the bottom quark into a larger representation, which will generate a term with the insertion of two $\Sigma$ fields.

\subsection{Antisymmetric tensor representation $\mathbf{15}$ and $\mathbf{\bar{15}}$}

The minimal choice is to have a bit larger $|\Delta_b|=3/2$. There are two cases, case (1) with $\Delta_b=3/2$ and case (2) with $\Delta_b=-3/2$. By analyzing the quantum numbers, we expect to generate bottom Yukawa coupling terms as
\begin{align}
(1)~\bar{q}_Ls{H}_1b_R\quad\text{and}\quad(2)~\bar{q}_Ls^*\tilde{H}_2b_R.
\label{expected_terms}
\end{align}

To realize such $|\Delta_b|$, the minimal choice is to use antisymmetric tensor representation $\mathbf{15}$ and $\mathbf{\bar{15}}$. To mix the SM quarks with composite operators, we first analyze their SM quantum numbers. To have operators sharing the same quantum numbers with the SM quarks, additional gauge $U(1)_X$ and global $U(1)_R$ are required. With additional $x$ and $r$ charges, the representation $\mathbf{15}_{x,r}$ can be decomposed under $SU(2)_W\times U(1)_Y\times U(1)_{F}$ as
\begin{align}
\mathbf{15}_{x,r}=&
\mathbf{(3\oplus1)}_{x,r}\oplus\mathbf{2}_{x+\frac{1}{2},r+\frac{1}{2}}\oplus\mathbf{2}_{x-\frac{1}{2},r+\frac{1}{2}}\oplus\mathbf{\bar{2}}_{x+\frac{1}{2},r-\frac{1}{2}}\nonumber\\
&\oplus\mathbf{\bar{2}}_{x-\frac{1}{2},r-\frac{1}{2}}\oplus\mathbf{1}_{x,r+1}\oplus\mathbf{1}_{x,r}\oplus\mathbf{1}_{x,r-1},
\end{align}
where the first subscript denotes its hypercharge and the second subscript denotes its flavor charge. Or we can write them in matrix form as
\begin{align}
&\mathbf{15}_{x,r}=
\begin{pmatrix}
1_{x,r+1}   &  2_{x+\frac{1}{2},r+\frac{1}{2}}  &  (3\oplus1)_{x,r}  &  2_{x-\frac{1}{2},r+\frac{1}{2}} \\
\cdot   &  0  &  \bar{2}_{x+\frac{1}{2},r-\frac{1}{2}}  &  1_{x,r}\\
\cdot   &  \cdot  &  1_{x,r-1} &  \bar{2}_{x-\frac{1}{2},r-\frac{1}{2}}\\
\cdot   &  \cdot  &  \cdot  &  0  \\  
\end{pmatrix},
\end{align}
and also for it complex conjugate $\mathbf{\bar{15}}_{x,r}$ as
\begin{align}
&\mathbf{\bar{15}}_{x,r}=
\begin{pmatrix}
1_{x,r-1}   &  \bar{2}_{x-\frac{1}{2},r-\frac{1}{2}}  &  (3\oplus1)_{x,r}  &  \bar{2}_{x+\frac{1}{2},r-\frac{1}{2}} \\
\cdot   &  0  &  {2}_{x-\frac{1}{2},r+\frac{1}{2}}  &  1_{x,r}\\
\cdot   &  \cdot  &  1_{x,r+1} &  {2}_{x+\frac{1}{2},r+\frac{1}{2}}\\
\cdot   &  \cdot  &  \cdot  &  0  \\  
\end{pmatrix}.
\end{align}
Since they are antisymmetric, we only put the numbers on the up-right triangle for simplicity.

\subsection{Two ways to embed the bottom quark}

Next, we want to mix the left-handed bottom quark with $\mathbf{15}$ and the right-handed bottom quark with $\mathbf{\bar{15}}$. The goal is to find a pair with $|\Delta_b|=3/2$. From the previous decomposition, we found two pairs that satisfy our requirement: 
\begin{equation}
\left(2_{x+\frac{1}{2},r+\frac{1}{2}},1_{x,r-1}\right) \quad\text{and}\quad 
\left(\bar{2}_{x+\frac{1}{2},r-\frac{1}{2}},1_{x,r+1}\right),\nonumber
\end{equation}
which correspond to case (1) with $\Delta_b=3/2$ and case (2) with $\Delta_b=-3/2$ respectively.

Let us start with case (1) by taking the first pair with $x=-1/3$ and $r=0$. Just as we have done before, we first write down the composite operators and the corresponding composite resonances as
\begin{equation}
O_{L,R}^{ij}\sim{\xi^i}_\alpha {\xi^j}_\beta Q_{L,R}^{\alpha\beta}\sim
\mathbf{15}_{-\frac{1}{3},0}=\mathbf{14}_{-\frac{1}{3},0}\oplus\mathbf{1}_{-\frac{1}{3},0},
\end{equation}
where $Q_{L,R}$ are the corresponding composite resonances. $Q_{L,R}$ are $\mathbf{14}$ and $\mathbf{1}$ of $Sp(6)$ and play the roles of the SM fermion composite partners.

The mixing term for the left-handed quark can be expressed as
\begin{equation}
\lambda_{b_L}\bar{q}_{La}{\Lambda^a}_{ij}O_R^{ij}=
\lambda_{b_L}\bar{q}_{La}{\Lambda^a}_{ij}\left({\xi^i}_\alpha {\xi^j}_\beta Q_{L,R}^{\alpha\beta}\right),
\end{equation}
where
\begin{equation}
{(\Lambda)^a}_{ij}=(
\begin{pmatrix}
0   &  0  &  1  & 0  & 0  & 0  \\
0   &  0  &  0  & 0  & 0  & 0  \\
-1   &  0  &  0  & 0  & 0  & 0  \\
0   &  0  &  0  & 0  & 0  & 0  \\
0   &  0  &  0  & 0  & 0  & 0  \\
0   &  0  &  0  & 0  & 0  & 0  \\
\end{pmatrix},
\begin{pmatrix}
0   &  0  &  0  & 0  & 0  & 0  \\
0   &  0  &  1  & 0  & 0  & 0  \\
0   &  -1  &  0  & 0  & 0  & 0  \\
0   &  0  &  0  & 0  & 0  & 0  \\
0   &  0  &  0  & 0  & 0  & 0  \\
0   &  0  &  0  & 0  & 0  & 0  \\
\end{pmatrix})
\end{equation}
is the spurion that can help us keep track of symmetry breaking.

We still need to mix the right-handed bottom quark with the composite operators and the corresponding composite resonances as $
O'_{L,Rij}\sim{{\xi^*}_i}^\alpha {{\xi^*}_j}^\beta \Sigma_{0\alpha\rho}\Sigma_{0\beta\sigma}Q_{L,R}^{\rho\sigma}\sim \mathbf{\bar{15}}_{-\frac{1}{3},0}.
$
The right-handed bottom quark need to mix with the $\mathbf{1}_{-\frac{1}{3},-1}$ of the $\mathbf{\bar{15}}_{-\frac{1}{3},0}$, which can be written as
\begin{equation}
\lambda_{b_R}\bar{b}_{R}{\Gamma}^{ij}O'_{Lij}=
\lambda_{b_R}\bar{b}_{R}{\Gamma}^{ij}\left({\xi^*_i}^\alpha {\xi^*_j}^\beta \Sigma_{0\alpha\rho}\Sigma_{0\beta\sigma}Q_{L,R}^{\rho\sigma} \right),
\end{equation}
where
\begin{equation}
{(\Gamma)}^{ij}=
\begin{pmatrix}
0   &  1  &  0  & 0  & 0  & 0  \\
-1   &  0  &  0  & 0  & 0  & 0  \\
0   &  0  &  0  & 0  & 0  & 0  \\
0   &  0  &  0  & 0  & 0  & 0  \\
0   &  0  &  0  & 0  & 0  & 0  \\
0   &  0  &  0  & 0  & 0  & 0  \\
\end{pmatrix}
\end{equation}
 is the corresponding spurion.

Combining $\lambda_{b_L}$ and $\lambda_{b_R}$ couplings, we can generate the bottom quark Yukawa coupling as
\begin{align}
&\sim
\lambda_{b_L}\lambda_{b_R}\bar{q}_{La}{\Lambda^a}_{ij}{\xi^i}_\alpha {\xi^j}_\beta \Sigma_{0}^{\alpha\rho}\Sigma_{0}^{\beta\sigma}{{\xi^T}_\rho}^k{{\xi^T}_\sigma}^l
{{\Gamma}^\dagger}_{kl} b_R\nonumber\\
&=\lambda_{b_L}\lambda_{b_R}\bar{q}_{La}{\Lambda^a}_{ij} \Sigma^{ik}\Sigma^{jl}{{\Gamma}^\dagger}_{kl} b_R
\supset \lambda_{b_L}\lambda_{b_R}\left(\bar{q}_Ls{H}_1b_R\right)~,
\label{eq:bottom_yukawa_1}
\end{align}
which is exactly what we expect in Eq.~\eqref{expected_terms}. The bottom quark gets mass from $H_1$ but with additional suppression from the FN mechanism as
\begin{align}
m_b=\frac{\langle is\rangle}{f}\frac{\lambda_{b_L}\lambda_{b_R}}{g_B}\frac{v_1}{\sqrt{2}}
=\frac{\lambda_{b_L}\lambda_{b_R}}{g_B}\frac{v_sv_1}{{2f}},
\end{align}
where $g_B$ is a coupling of the composite bottom partners. This is like a Type-II 2HDM but with smaller tan$\beta$ due to the suppression by small $v_s/f$.

Therefore, for case (1), we can get the top-bottom mass hierarchy. Assuming all the $\lambda$ and $g$ are $\mathcal{O}(1)$ couplings, the mass ratio bocomes \footnote{We consider the running of quark masses up to 1 TeV \cite{Xing:2007fb}, which gives $m_b/m_t\sim 2.43 \text{ GeV}/150 \text{ GeV}\sim 1/60$. The ratio might be larger because the VEV of the flavon field is below the TeV scale.} 
\begin{align}
\frac{m_b}{m_t}\sim \frac{v_s}{\sqrt{2}f}\frac{v_1}{v_2}=\frac{\epsilon}{\text{tan}\beta} \sim\frac{1}{60}~,
\end{align}
where $\epsilon\equiv{v_s}/{\sqrt{2}f}$. The hierarchy comes from both $\epsilon$ and tan$\beta$. Taking the symmetry breaking scale $f \sim 1$ TeV, we get
\begin{align}
v_s \sim 25~\text{tan}\beta \text{ GeV,}
\end{align}
If $\epsilon$ (namely $v_s$) is small, we can get a Type-II 2HDM with a smaller tan$\beta$.

Similarly, consider case (2) by taking the second pair with $x=-1/3$ and $r=1$, i.e. $\left(\bar{2}_{\frac{1}{6},\frac{1}{2}},1_{-\frac{1}{3},2}\right)$. The spurion for the left-handed quark becomes
\begin{equation}
{(\Lambda)^a}_{ij}=(
\begin{pmatrix}
0   &  0  &  0  & 0  & 0  & 0  \\
0   &  0  &  0  & 0  & 0  & 0  \\
0   &  0  &  0  & 0  & 1  & 0  \\
0   &  0  &  0  & 0  & 0  & 0  \\
0   &  0  &  -1  & 0  & 0  & 0  \\
0   &  0  &  0  & 0  & 0  & 0  \\
\end{pmatrix},
\begin{pmatrix}
0   &  0  &  0  & 0  & 0  & 0  \\
0   &  0  &  0  & 0  & 0  & 0  \\
0   &  0  &  0  & 1  & 0  & 0  \\
0   &  0  &  -1  & 0  & 0  & 0  \\
0   &  0  &  0  & 0  & 0  & 0  \\
0   &  0  &  0  & 0  & 0  & 0  \\
\end{pmatrix})
,
\end{equation}
and for the right-handed bottom quark is
\begin{equation}
{(\Gamma)}^{ij}=
\begin{pmatrix}
0   &  0  &  0  & 0  & 0  & 0  \\
0   &  0  &  0  & 0  & 0  & 0  \\
0   &  0  &  0  & 0  & 0  & 0  \\
0   &  0  &  0  & 0  & 1  & 0  \\
0   &  0  &  0  & -1  & 0  & 0  \\
0   &  0  &  0  & 0  & 0  & 0  \\
\end{pmatrix}~.
\end{equation}

Combining $\lambda_{b_L}$ and $\lambda_{b_R}$ couplings in case (2), we get the bottom Yukawa coupling as
\begin{align}
&\sim
\lambda_{b_L}\lambda_{b_R}\bar{q}_{La}{\Lambda^a}_{ij}{\xi^i}_\alpha {\xi^j}_\beta \Sigma_{0}^{\alpha\rho}\Sigma_{0}^{\beta\sigma}{{\xi^T}_\rho}^k{{\xi^T}_\sigma}^l
{{\Gamma}^\dagger}_{kl} b_R\nonumber\\
&=\lambda_{b_L}\lambda_{b_R}\bar{q}_{La}{\Lambda^a}_{ij} \Sigma^{ik}\Sigma^{jl}{{\Gamma}^\dagger}_{kl} b_R
\supset \lambda_{b_L}\lambda_{b_R}\left(\bar{q}_Ls^*\tilde{H}_2b_R\right)~.
\label{eq:bottom_yukawa_2}
\end{align}
Again it is what we expect in Eq.~\eqref{expected_terms}. This case will lead to a Type-I 2HDM with the small bottom Yukawa coupling merely due to the FN mechanism as
\begin{align}
m_b=\frac{\langle is\rangle}{f}\frac{\lambda_{b_L}\lambda_{b_R}}{g_B}\frac{v_2}{\sqrt{2}}
=\frac{\lambda_{b_L}\lambda_{b_R}}{g_B}\frac{v_sv_2}{{2f}}.
\end{align}
Assuming all the $\lambda$ are $\mathcal{O}(1)$ couplings. The mass ratio
\begin{align}
\frac{m_b}{m_t}\sim \frac{v_s}{\sqrt{2}f}={\epsilon} \sim\frac{1}{60}
\quad\implies\quad 
v_s \sim 25 \text{ GeV, }
\end{align}
if the symmetry breaking scale $f \sim 1$ TeV.

\subsection{Composite resonances and spaghetti diagrams}

In the last section, we see how the FN mechanism can be realized and create the hierarchy between the top and bottom mass. The composite resonances, which carry the same quantum number but different flavor charges, play the role of the Froggatt-Nielsen fields in the FN mechanism. We can write down all the composite resonances in matrix form as
\begin{align}
\mathbf{15}_{-\frac{1}{3},0}&=\mathbf{14}_{-\frac{1}{3},0}\oplus\mathbf{1}_{-\frac{1}{3},0} \nonumber\\
&=
\begin{pmatrix}
0   &  \tilde{B}_1  &  T_\frac{1}{2}  & \tilde{B}'_0  & \tilde{T}_0  & \tilde{B}_\frac{1}{2}  \\
\cdot   &  0  &  B_\frac{1}{2}  & Y_0  & \tilde{B}''_0  & Y_\frac{1}{2}  \\
\cdot   &  \cdot  &  0  & B_{-\frac{1}{2}}  & T_{-\frac{1}{2}}  & 0  \\
\cdot   &  \cdot  &  \cdot  & 0  & \tilde{B}_{-1}  & Y_{-\frac{1}{2}}  \\
\cdot   &  \cdot  &  \cdot  & \cdot  & 0  & \tilde{B}_{-\frac{1}{2}}  \\
\cdot   &  \cdot  &  \cdot  & \cdot  & \cdot  & 0  \\
\end{pmatrix}
\oplus \tilde{B}_0,
\end{align}
where $T$ and $B$ are composite resonances with the same quantum numbers as the SM top and bottom quarks but with different flavor charges as labeled in the subscript, $\tilde{T}$ and $\tilde{B}$ are resonances with the same hypercharges as the SM top and bottom quarks but under different $SU(2)_W$ representations, and $Y$ are exotic resonances with hypercharge $-4/3$.

\begin{figure}[t]
\centering
\includegraphics[width=1.0\linewidth]{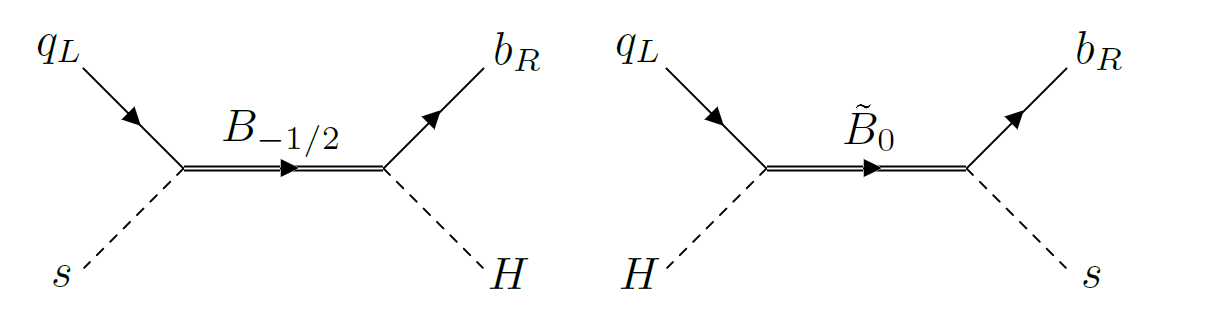}
\caption{Spaghetti diagrams that generate the bottom Yukawa coupling through the Froggatt-Nielsen mechanism in case (1). Diagrams for case (2) are similar.}
\label{bottom}
\end{figure}

The FN mechanism can also be expressed through the ``spaghetti diagrams'', which looks like a 2 to 2 scattering in this case with only one flavon inserted. Spaghetti diagrams that generate the suppressed bottom quark mass are shown in Fig.~\ref{bottom}. These diagrams give us the bottom mass we expect after integrating out the heavy Froggatt-Nielsen fields, which are composite fermionic resonances in this model, and replacing the scalar fields with their VEVs.

\subsection{Comparison between two cases}

So far, we see two different flavor charge assignments for the right-handed bottom quark, which lead to two different bottom Yukawa coupling terms. Both of them successfully generate a suppressed bottom Yukawa coupling through the FN mechanism. The difference between these two cases is listed in Table ~\ref{comparison}.
\begin{table}[tb]
\begin{tabular}{|c|c|c|c|}
\hline
& Case~(0) & Case~(1) & Case~(2) \\ \hline
$\Delta_b \equiv [q_L]-[b_R]$ & $1/2$ & $3/2$ & $-3/2$ \\ \hline
Flavor charge of $b_R$ & 0 & -1 & 2 \\ \hline
Coupling term & $\bar{q}_L{H}_1b_R$ & $\bar{q}_Ls {H}_1b_R$ & $\bar{q}_Ls^*\tilde{H}_2b_R$ \\ \hline
Type of 2HDM & Type-II & Type-II & Type-I \\ \hline
Suppression of $m_b/m_t$ & $1/{\text{tan}\beta}$ & ${\epsilon}/{\text{tan}\beta}$ & ${\epsilon}$ \\ \hline
\end{tabular}
\caption{The comparison between two cases with suppressed bottom Yukawa couplings through the FN mechanism. Case (0) for the unsuccessful first taste is also shown. In the last row, $\epsilon\equiv{v_s}/{\sqrt{2}f}$ is the suppression by the FN mechanism. \label{comparison}}
\end{table}

Here we assume the flavor charge of $q_L$ is 1/2 and $t_R$ is 0, such that the top quark mass comes from $H_2$. We can see the two cases represent different signs of $\Delta_b$. It will affect the way we extend our model to include lighter quarks, which will be discussed next. The difference between the types of 2HDM results in different Higgs phenomenology. The second Higgs doublet is expected to be the main target among the exotic states in the model. The results of the direct searches will be shown in the next section. The factor of suppression is also related to the experimental constraint. The smaller $\epsilon$ required for the correct mass ratio implies a smaller VEV $v_s$ of the flavon field, which will end up with a larger deviation in flavor observables and thus is strongly constrained.

\subsection{Include all the generations}

As yet, we only get the hierarchy between the top and bottom quarks, which belong to the third generation. To include the lighter quarks, more suppression is needed, which means more insertion of the flavon field $s$ and a larger difference in flavor charges. This will require the lighter quarks to be embedded in even larger representations.

Take case (1) for example. We have already gotten the flavor charges of the third generation quarks. To extend to the first and the second generations, one possible flavor charges assignment \footnote{The assignment is borrowed from Eq. (2.15) of \cite{Bauer:2015fxa}. The resulting mass ratios and CKM matrix can partially reproduce observed values with $\mathcal{O}(1)$ correction.} is listed in Table~\ref{tab:FN1}.
It implies that we need even larger representations to have flavor charges different by $7/2$. That would require representations with more than 4 indices for the quark sector.

\begin{table}[b]
\begin{tabular}{||c|c||c|c||c|c||}
\hline
& $U(1)_{F}$ & & $U(1)_{F}$ & & $U(1)_{F}$ \\ \hline
$q_{3,L}=(t_L,b_L)^T$ & 1/2 & $t_R$ & 0 & $b_R$ & -1 \\ \hline
$q_{2,L}=(c_L,s_L)^T$ & 3/2 & $c_R$ & 0 & $s_R$ & -1 \\ \hline
$q_{1,L}=(u_L,d_L)^T$ & 3/2 & $u_R$ & -2 & $d_R$ & -2 \\ \hline
\end{tabular}
\caption{A possible flavor charge assignment of all elementary quarks for case (1) setup.\label{tab:FN1}}
\end{table}

For case (2), it is more difficult to get a consistent flavor charges assignment for the desired CKM matrix. For the flavor charge of the third generation quarks, we find that they follow the order $[b_R]>[q_L]>[t_R]$, which is also applied to the extension. From the relation, the left-handed quarks should always sit in the middle. This requirement restricts the flavor charge difference we can have. For example, $q_{1,L}$ and $q_{2,L}$ can only be either $1/2$ or $3/2$, which will lead to unsuppressed entries in the CKM matrix. Therefore, considering the flavor charge assignment for the light quarks, case (1) is preferred over case (2). However, we will still discuss the constraints on parameter space of case (2) assuming that it can generate a similar Yukawa matrix as case (1).

The exact embedding will be explored in future work. To discuss the experimental constraints of flavons in the following section, we will assume this mechanism can be extended to all the generations and is responsible for all the light quark masses in both cases. Also, for flavon phenomenology, the results are mainly determined by two parameters, the flavon mass $M_s$ and the flavon VEV $v_s$.

\section{Collider Signature}\label{sec:Collider}

The phenomenology of this model is similar to other CHMs based on $SU(6)/Sp(6)$ coset with partial compositeness \cite{Cheng:2020dum}, which includes 14 pNGBs and composite partners of the SM particles. The main targets would be on the particles that couple to SM particles at leading order. In our setup, the most important search modes include the second Higgs doublet, flavons, and fermionic composite resonances.

\subsection{The second Higgs Doublet}

The phenomenology of 2HDM has been well-studied, and we can directly borrow the results from \cite{Kling:2020hmi}. For case (2) as a Type-I 2HDM, there is no further constraint since the second Higgs doublet is decoupled from the fermion sector. But for case (1), a Type-II 2HDM, the constraints are important because the suppression of the bottom mass comes partially from the FN mechanism and partially from tan$\beta$. Therefore, the value of tan$\beta$ will decide the $\epsilon$ we need from the FN mechanism. The strongest constraint for a Type-II 2HDM comes from the $\tau\tau$ search, which restricts tan$\beta<6-10$ for a wide mass scale. If we make it a Flipped 2HDM instead, where the charged leptons get masses from $H_2$ instead of $H_1$, the coupling between heavy Higgs and $\tau\tau$ will become much smaller. Then the main constraint comes from $\bar{b}b$ search, and tan$\beta\sim20$ is still allowed. However, we would like to stick to a normal Type-II 2HDM for case (1) and set the benchmark with
\begin{align}
\text{tan}\beta\sim6
\quad\implies\quad
v_s \sim 150 \text{ GeV}
\end{align}
for the following discussion.

\subsection{Flavons}

The physical flavon fields include a scalar component $s$ and a pseudoscalar component $a$. The masses of two types of flavons depend on the complete flavon potential, which is discussed in the appendix \ref{Potential}. If flavor symmetry is exact and spontaneously broken by flavor symmetry conserving potential, then the pseudoscalar flavon should be massless, which is not acceptable. Therefore, the explicit breaking of flavor symmetry in the flavon potential is needed. For simplicity, we will assume the mass of scalar, $M_s$, and the mass of pseudoscalar, $M_a$, are the same. This spectrum can be achieved if flavor symmetry is broken by a tadpole term in the flavon potential as shown in appendix~\ref{Potential}. Therefore, from now on, we will use flavon $s$ for both the scalar and pseudoscalar components and $M_s$ for the flavon mass, which is expected to be at the sub-TeV scale.

The production and decay of flavons have already been comprehensively discussed in \cite{Tsumura:2009yf, Bauer:2016rxs}. Although the flavon coupling terms in these papers might look different from ours, the exact values are determined by the observed quark masses and the CKM matrix. Therefore, the flavon couplings with the form $m/v_s$ should have similar values in all kinds of flavon models up to an $\mathcal{O}(1)$ factor. The numerical values in these two sections are derived based on their analysis with additional adjustments from our setup.

The main production for sub-TeV flavons come from the single production process $b\bar{b}\to s$. The cross section for flavons with $M_s=500$ GeV is
\begin{align}
\sigma (b\bar{b}\to s)\sim 9.8\times10^{-3}\left(\frac{150 \text{ GeV}}{v_s}\right)^2\text{ pb}
\end{align}
in 14 TeV LHC. Taking $v_s=150$ GeV, around $2.2\times10^4$ flavons will be produced in the HL-LHC era with an integrated luminosity of 3 ab$^{-1}$. In case (2) with smaller $v_s$, the number is multiplied by a factor of 36.

The decay branching ratios for flavons are independent of $v_s$ but only depend on the flavor structure. If flavons only couple to the third generation, the dominate decay channel will be $b\bar{b}$ channel and $\tau\tau$ channel with roughly $\sim85\%$ and $\sim15\%$ branching ratio. If the FN mechanism is extended to all SM particles and responsible for the full Yukawa matrix, then there will be exotic final states like $tc$ and $tu$. It turns out that the $tc$ channel will be the dominant one due to the large mixing required to reproduce the desired CKM matrix. The ratios depend on tan$\beta$, too. Under the benchmark values, we get the branching ratios for each channels as $tc~(96.8\%)$, $b\bar{b}~(2.7\%)$, and $\tau\tau~(0.5\%)$. However, the hadronic channels suffer from large backgrounds. The leptonic channel can reach $\sigma\times BR\sim 10^{-3}$ pb for sub-TeV flavon in HL-LHC, but it is still above the benchmark value. The discovery can be made in a future 100 TeV collider, where the cross section is expected to be $\sim100$ times larger, and the integrated luminosity is also higher. In that case, the distinct $tc$ channel search will provide strong evidence for the origin of the Yukawa matrix.

\subsection{Fermionic Resonances}

The top partners in the $SU(6)/Sp(6)$ CHM are vector-like fermionic resonances that form a sextet of the $Sp(6)$ global symmetry. Their quantum numbers under the SM gauge symmetry are $(3,2,1/6) [\times 2]$, $(3,1,2/3)$, and $(3,1,-1/3)$, which are identical to those of the SM quarks. There are no exotic states with higher or lower hypercharges. These states are degenerate in the limit of unbroken $Sp(6)$ global symmetry. Only small splittings arise from the explicit symmetry breaking effects. Their mass $M_T\sim g_Tf$ plays the important role of cutting off the quadratic contribution from the top quark loop to the Higgs potential. The generic expectation of the composite fermionic resonances is $M_F=5-10$ TeV with $g_F=5-10$. However, naturalness prefers a smaller $M_T$ to minimize the required fine-tuning, which usually requires $g_T\gtrsim1$. The current bound on the top partner mass has reached $\sim 1.2$~TeV~\cite{Sirunyan:2018omb, Aaboud:2018pii}. The HL-LHC can further constrain the mass up to $\sim 1.5$~TeV~\cite{CidVidal:2018eel}. A future 100 TeV collider will cover the entire interesting mass range of the top partners if no severe tuning conspires.

For the bottom partners, they form a $\mathbf{14}_{-1/3}\oplus\mathbf{1}_{-1/3}$ under $Sp(6)$ global symmetry. The quantum numbers for the total of 15 fields under the SM gauge group are
$(3,2,1/6) [\times 2]$, $(3,2,-5/6) [\times 2]$, $(3,1,-1/3) [\times 4]$, and $(3,3,-1/3)$, which include exotic resonances with EM charge $-4/3$. The states are not degenerate, and the singlet is expected to be lighter. The masses of the bottom partners $M_B\sim g_Bf$, unlike the top partners, don't have a large effect on the fine-tuning due to the small bottom Yukawa coupling. Therefore, they could be around the compositeness scale with $M_B=5-10$ TeV, which is beyond the LHC searches. The heavier $M_B\sim g_Bf$ also leads to additional suppression $g_T/g_B$ on the mass ratio between the top and bottom quarks, which can relieve the required $\epsilon$ we need.

If we extend the FN mechanism to the light generations, a larger representation is required to get a larger flavor charge difference, which also implies a larger EM charge difference within the multiplet. Therefore, there could be more exotic resonances with EM charges like $-7/3$ or $5/3$, which are important in identifying the correct representation. These heavy fermionic resonances can be found in a future 100 TeV collider. If the exotic spectrum corresponding to the large representation shows up, it might unveil the nature of SM fermion partners and the origin of Yukawa couplings.

\section{Flavor constraints}\label{sec:Flavor}

Compared to the collider signatures, the flavor constraints usually probe a higher scale and place stronger bounds on the models. Assume that the FN mechanism can be extended to all elementary quarks and leptons with suitable Yukawa coupling matrices. Then we can discuss the flavor constraints through a similar analysis as in \cite{Bauer:2016rxs}.

The new flavor processes can be mediated through flavons or the second Higgs doublet. The flavon contributions strongly depend on the couplings and spectrum of flavons. As we mention above, there are a scalar component and a pseudoscalar component. We will assume the scalar and pseudoscalar components share the same mass $M_s$. This assumption will give us the weakest flavor constraints because, for some flavor processes, the contributions from a scalar and a pseudoscalar will cancel exactly if they are degenerate. It can also be understood that the assumption raises an $U(1)$ symmetry for the flavon field around the vacuum, which forbids these flavor processes. However, we will see even the weakest constraints from flavor are much stronger than the direct searches.

\subsection{Meson Decay}

The new particles might enhance some rare processes that are suppressed within SM. The measurements of rare decays of neutral mesons can give strong constraints on the new physics scale. In this model, flavons can mediate some rare decays of neutral mesons. For example, the branching ratio of $B_s\to \mu^+ \mu^-$ provides a constraint on dimension-6 operators induced by flavons, which include
\begin{align}
C_S^{ij}(\bar{q}_iP_Lq_j)(\bar{\ell}\ell) \quad\text{and}\quad \tilde{C}_S^{ij}(\bar{q}_iP_Rq_j)(\bar{\ell}\ell)
\end{align}
from a scalar flavon with coefficient
\begin{align}
C_S^{ij}=g_{\ell\ell}g_{ji}\left(\frac{1}{M_s^2}\right) \quad\text{and}\quad
\tilde{C}_S^{ij}=g_{\ell\ell}g_{ij}\left(\frac{1}{M_s^2}\right)
\end{align}
and
\begin{align}
C_P^{ij}(\bar{q}_iP_Lq_j)(\bar{\ell}\gamma_5\ell) \quad\text{and}\quad \tilde{C}_P^{ij}(\bar{q}_iP_Rq_j)(\bar{\ell}\gamma_5\ell)
\end{align}
from a pseudoscalar flavon with coefficient
\begin{align}
C_P^{ij}=g_{\ell\ell}g_{ji}\left(\frac{1}{M_a^2}\right) \quad\text{and}\quad
\tilde{C}_P^{ij}=g_{\ell\ell}g_{ij}\left(\frac{1}{M_a^2}\right).
\end{align}
The difference between $C$ and $\tilde{C}$ will modify the predicted SM values. The leading order deviation comes from the pseudoscalar flavon exchange, which interferes with the SM contribution. The coupling $g_{ij}$ is determined by the observed fermion masses over the flavon VEV $v_s$. Therefore, once we take the mass $M_s=M_a$, the measurement can put a constraint on the $C_P^{ij}-\tilde{C}_P^{ij}$ and thus the product of $v_sM_s$. Later we will find that most of the flavor constraints can be transferred into the constraint on the value of $v_sM_s$.

The latest result of $B_s\to\mu^+ \mu^-$ measurement by LHCb \cite{LHCb2021} requires $v_sM_s\geq 5\times 10^4 \text{ (GeV)}^{2}$, which give a $M_s$ lower bound under the benchmark value as
\begin{align}
\text{case (1) }M_s\geq 400 \text{ (GeV)},\quad \text{case (2) }M_s\geq 2000 \text{ (GeV)}.\nonumber
\end{align}
There is a stronger constraint for case (2) flavon model with smaller $v_s$. The reason is, though we want to have a small $v_s$ to generate the hierarchy, a small $v_s$ also implies a larger coupling between flavons and the SM quarks, which is disfavored by flavor physics. We also find that case (1) as a Type-II 2HDM has a looser bound due to the assistance from tan$\beta$. The improvement in the measurement of BR($B_s\to\mu^+ \mu^-$) will further constraints the allowed values in the future. The interesting parameter space might be ruled out by LHCb and Belle-II.

Meson decays also put strong constraints on the second Higgs doublet. A light charged Higgs boson can induce a significant contribution to the branching ratio BR$(B\to X_s\gamma)$~\cite{Grinstein:1987pu, Hou:1987kf, Rizzo:1987km, Geng:1988bq, Barger:1989fj, Grinstein:1990tj}. In the Type-II or flipped 2HDM, this gives a strong lower bound on the charged Higgs boson mass $M_{H^\pm}>600$ GeV~\cite{Arbey:2017gmh, Misiak:2017bgg}, which would require a tuning or an additional symmetry in the 2HDM potential in case (1) model.

\subsection{Neutral Meson Mixing}

The strongest bounds for flavons come from the neutral meson mixing, especially from the light mesons. The relevant $\Delta F=2$ interaction terms include
\begin{align}
C_2^{ij}(\bar{q}^i_Rq^j_L)^2,\quad \tilde{C}_2^{ij}(\bar{q}^i_Lq^j_R)^2,\quad\text{and}\quad C_4^{ij}(\bar{q}^i_Rq^j_L)(\bar{q}^i_Lq^j_R).\nonumber
\end{align}

In this paper, since we assume that the scalar and pseudoscalar flavons share the same mass $M_s$, there is an $U(1)$ symmetry that forbids $C_2^{ij}$ and $\tilde{C}_2^{ij}$ terms. That is, the contributions from scalars and pseudoscalars will cancel exactly. The only relevant dimension-6 operator is
\begin{align}
C_4^{ij}(\bar{q}^i_Rq^j_L)(\bar{q}^i_Lq^j_R)
\quad\text{ with}\quad C_4^{ij}=-g_{ij}g^*_{ji}\left(\frac{1}{M_s^2}\right).
\end{align}
The coefficients as a function of $v_sM_s$ are strongly constrained by experiments.

In Table \ref{MesonMixing}, we conclude the flavor constraints on the product $v_sM_s$ from all neutral meson systems, including those with the first generation quarks. The numbers are extracted from \cite{Bauer:2016rxs}. The corresponding lower bounds on flavon mass $M_s$ are also shown based on the benchmark value of each case.

\begin{table}[tb]
\begin{tabular}{|c|c|c|c|}
\hline
 & $v_sM_s$ (GeV$^2$) & Case (1) (GeV)  & Case (2) (GeV) \\ \hline
$	C_{B_s}	$	&	32000	&	210	&	1280	\\	\hline
$	\varphi_{B_s}	$	&	128000	&	850	&	5120	\\	\hline
$	C_{B_d}	$	&	183000	&	1220	&	7320	\\	\hline
$	\varphi_{B_d}	$	&	250000	&	1670	&	10000	\\	\hline
$	\Delta m_K	$	&	255000	&	1700	&	10200	\\	\hline
$	\epsilon_K	$	&	2550000	&	17000	&	102000	\\	\hline
\end{tabular}
\caption{Flavor constraints from all kinds of neutral meson mixing observables, including the lower bounds on the value of $v_sM_s$ and flavon mass $M_s$ of each case. \label{MesonMixing}}
\end{table}

From the constraints of neutral meson mixing, we again find that case (1) is preferred because case (2) has a smaller $v_s$ and thus larger couplings to the SM fermions. The lower bounds for case (2) have reached multi-TeV, which might be too heavy to be treated as pNGBs. The flavor symmetry is hardly broken, and the sigma model might not be an appropriate way to describe it. Even for case (1) with milder bounds, constraints from the CP phases are also high. If we assume that the flavon preserves CP-symmetry and ignore the constraints from the CP phase, the current bounds for case (1) become $M_s\geq 1.2-1.7$ TeV, and the future experiments will raise the bounds by a factor of 2. If the FN mechanism is not responsible for the first generation quarks, then the only constraint is from $C_{B_s}$, and a sub-TeV flavon is still allowed. The bounds can also be relieved if the bottom partner is heavier than the top partner, where $g_B>g_T$ can give another suppression, and the required $v_s$ can be larger. Nevertheless, the most interesting mass region for flavons as pNGBs of the TeV scale confinement will be covered in the near future by LHCb and Belle-II.

\section{Conclusions}\label{sec:Conclusion}

The Froggatt-Nielsen mechanism is an appealing solution to the Flavor Puzzle. However, the scale of flavor dynamics and the flavon field can be arbitrarily high. The predictive flavon models require the dynamics to stabilize the flavon potential. One way, analogous to the composite Higgs models, is to introduce the flavon field as a pseudo-Nambu-Goldstone boson. In this paper, we construct a non-linear sigma model with pNGBs, including both the Higgs doublets and the flavon field.

The flavon field as a pNGB provides a possibility to have the origin of flavor hierarchy at the TeV scale. The shift symmetry is slightly broken, which leads to the flavon mass and VEV. The non-linear nature of the flavon also constraints the interactions we can write down. In this paper, we show two possible ways to generate suppressed bottom Yukawa coupling terms through the Froggatt-Nielsen mechanism, where the composite resonances play the role of the FN fields. The derivation and explanation of the process are presented in detail.

Two cases lead to different phenomenology and receive different constraints. Case (1) as a Type-II 2HDM with small tan$\beta$ has a larger $v_s$ and smaller couplings to the SM fermions. Some parameter space with the sub-TeV flavon is still allowed if the constraints from the neutral meson of the first generation quarks are not taken into account. Case (2) as a Type-I 2HDM has a weaker bound on the Higgs sector. However, the requirements of small $v_s$ and the strong couplings with the SM particles are disfavored. Future measurements of neutral meson systems by LHCb and Belle-II will keep probing the scenario with the light flavon. Either push the mass bound to a much higher scale or find the existence of the pNGB flavon.

\acknowledgments

We thank Hsin-Chia Cheng for useful discussions. This work is supported by the Department of Energy Grant number DE-SC-0009999.

\appendix

\section{The SM gauge sector}\label{Gauge}

The SM electroweak gauge group $SU(2)_W\times U(1)_Y$ is embedded in $SU(6)\times U(1)_X$ with generators given by
\begin{align}
\frac{1}{2}&
\begin{pmatrix}
\sigma^a   &  0  &  0  &  0 \\
0   &  0  &  0  &  0\\
0   &  0  &  -\sigma^{a*} &  0\\
0   &  0  &  0  &  0  \\  
\end{pmatrix}\text{ and }
\frac{1}{2}
\begin{pmatrix}
0_{2\times2}   &  0  &  0  &  0 \\
0   &  1  &  0  &  0 \\
0   &  0  &  0_{2\times2}  &  0 \\
0   &  0  &  0  &  -1  \\  
\end{pmatrix} + X \mathbf{I}~.\nonumber
\end{align}
The extra $U(1)_X$ factor accounts for the different hypercharges of the SM fermions but is not relevant for the bosonic fields. These generators belong to $Sp(6)\times U(1)_X$ and are not broken by $\Sigma_0$.

Using the $\Sigma$ field, the Lagrangian for kinetic terms of Higgs boson is given by
\begin{equation}
\mathcal{L}_h=\frac{f^2}{4}\text{tr}\left[(D_{\mu}\Sigma)(D^\mu \Sigma)^\dagger\right]+\cdots ,
\end{equation}
where $D_{\mu}$ is the electroweak covariant derivative. Expanding this term, we get
\begin{align}
\mathcal{L}_h&=\frac{1}{2}(\partial _\mu h_1)(\partial ^\mu h_1)+\frac{1}{2}(\partial _\mu h_2)(\partial ^\mu h_2)\nonumber\\
&+\frac{f^2}{2}g_W^2 \left(\text{sin}^2\frac{\sqrt{h_1^2+h_2^2}}{\sqrt{2}f}\right) \left[W^+_\mu W^{-\mu}+\frac{Z_\mu Z^\mu}{2\text{cos}\theta_W}\right]~.
\end{align}
The non-linear behavior of the Higgs boson in CHMs is apparent from the dependence of trigonometric functions.

The $W$ boson acquires a mass when $h_1$ and $h_2$ obtain nonzero VEVs $V_1$ and $V_2$ of
\begin{equation}
m_W^2=\frac{f^2}{2}g_W^2 \left(\text{sin}^2\frac{\sqrt{V_1^2+V_2^2}}{\sqrt{2}f}\right)= \frac{1}{4}g_W^2(v_1^2+v_2^2),
\end{equation}
where
\begin{equation}
v_i\equiv \sqrt{2}f\frac{V_i}{\sqrt{V_1^2+V_2^2}}\,\text{sin}\frac{\sqrt{V_1^2+V_2^2}}{\sqrt{2}f}\approx V_i=\langle h_i\rangle~.
\end{equation}
The parameter that parametrizes the nonlinearity of the CHM is given by
\begin{equation}
\xi\equiv \frac{v^2}{f^2}= 2\,\sin^2\frac{\sqrt{V_1^2+V_2^2}}{\sqrt{2}f}~,
\end{equation}
where the VEV $v^2=v_1^2+v_2^2=(246 \text{ GeV})^2$. The $\xi$ plays an important role in the phenomenology of CHMs, but it is not of interest in this study.

\section{The pNGB potential}\label{Potential}

The pNGB potential comes from the explicit breaking of $SU(6)$ global symmetry. Within SM, there are symmetry-breaking sources like the gauge couplings and Yukawa couplings. Additional sources are also needed to introduce the flavon potential. Here we will briefly list their contributions to the pNGB potential one by one.

Starting with the SM gauge interactions, we can derive the pNGB potential by the generators listed in appendix~\ref{Gauge}. Both $SU(2)_W$ and $U(1)_Y$ only break the global symmetry partially and generate the potential for the pNGBs which are charged. The two Higgs doublets are charged under both gauge interactions and get
\begin{align}
\Delta V_H=\frac{3}{16\pi^2}\left(\frac{3}{4}c_wg^2+\frac{1}{4}c'g'^2\right)M_\rho^2|H|^2,
\end{align}
where $M_\rho\sim g_\rho f$ is the mass of the vector resonances $\rho$, which act as the gauge boson partners to cut off the gauge loop contribution to the pNGB masses, and $c_w$ and $c'$ are $\mathcal{O}(1)$ constants. The scalar triplet $\phi$ also gets a potential
\begin{align}
\Delta V_\phi=\frac{3}{16\pi^2}\left(2c_wg^2\right)M_\rho^2(\phi^a\phi_a),
\end{align}
The $SU(2)_W \times U(1)_Y$ singlets $s$ and $\eta$ do not receive potentials from the gauge interactions at this order, but they will obtain potentials elsewhere.

Next, the Yukawa coupling also breaks the $SU(6)$ global symmetry. Take the top quark loop-induced potential for example, where the required spurions are already written in section~\ref{sec:Yukawa}. We can estimate
\begin{align}\label{H2_mass}
\Delta V_H&\sim -\frac{N_c}{8 \pi^2} \lambda_{L}^2\lambda_{R}^2 f^4 \left|{(\Lambda)^a}_i{(\Gamma^{*})}_j\Sigma^{ij}\right|^2\nonumber\\
&\supset-\frac{N_c}{8\pi^2}\lambda_{L}^2\lambda_{R}^2f^2\,|H|^2=-\frac{N_c}{8 \pi^2}y_t^2M_T^2\,|H|^2.
\end{align}
The dominant quartic term is also from the top loop as
\begin{align}
\Delta V_H\sim\frac{N_c}{4 \pi^2}y_t^4\,|H|^4.
\end{align}
Similar potentials also arise for other SM Yukawa interactions.

The real singlet $\eta$ does not get a potential at the leading order, but it couples quadratically to the Higgs doublets (e.g., from Eq.~\eqref{H2_mass}), so it can still obtain a potential after the Higgs doublets develop nonzero VEVs. Through Eq.~\eqref{H2_mass}, $\eta$ gets a quadratic potential
\begin{align}
\Delta V_\eta\sim \frac{3}{8\pi^2}y_t^2M_T^2\cdot \left(\frac{v}{f}\right)^2\eta^2.
\end{align}

So far, we haven't gotten any potential for the flavon field $s$. Although the flavon field in our model couples to the bottom quark, which will lead to a loop-induced pNGB potential. However, we would like to have the potential from a separate source, so they are independent of the FN mechanism. A nontrivial potential for the flavon field $s$ is common in models with collective symmetry breaking \cite{Low:2002ws, Cheng:2020dum}, where the potential
\begin{equation}
\Delta V\sim M_s^2\left|s\pm\frac{i}{2f}\tilde{H_2}^\dagger H_1\right|^2\supset M_s^2|s|^2
\end{equation}
is introduced. For example, in the $SU(6)/Sp(6)$ little Higgs model \cite{Low:2002ws}, the term can be generated by gauging two copies of $SU(2)$. However, it introduces new heavy gauge bosons $W'$ and $Z'$, which are strongly constrained.

Another way, following \cite{Cheng:2020dum}, is using the interactions between the elementary fermions and the resonances of the strong dynamics. In the section \ref{sec:Yukawa}, we see that the elementary quark doublets can couple to composite operators of $SU(6)$ representations $\mathbf{6}$ and $\mathbf{\bar{6}}$ with $x=1/6$ and $r=0$, which are decomposed under $SU(2)_W\times U(1)_Y\times U(1)_F$ as
\begin{subequations}
\begin{align}
&\mathbf{6}_{1/6,0}=\mathbf{2}_{1/6,1/2}\oplus\mathbf{1}_{2/3,0}\oplus\mathbf{\bar{2}}_{1/6,-1/2}\oplus\mathbf{1}_{-1/3,0},\\
&\mathbf{\bar{6}}_{1/6,0}=\mathbf{\bar{2}}_{1/6,-1/2}\oplus\mathbf{1}_{-1/3,0}\oplus\mathbf{2}_{1/6,1/2}\oplus\mathbf{1}_{2/3,0}.
\end{align}
\end{subequations}
Both operators create the same resonances, which belong to $\mathbf{6}$ of the $Sp(6)$ group.

Now consider two elementary quark doublets, $q_L$ and $q'_L$, couple to the first two components of the composite operators of $\mathbf{6}$ and $\mathbf{\bar{6}}$ respectively, while both representations contain the same resonances:
\begin{equation}
\lambda_{L}\bar{q}_{La}{\Lambda^a}_iO_R^i=
\lambda_{L}\bar{q}_{La}{\Lambda^a}_i\left({\xi^i}_\alpha Q_{R}^\alpha \right),
\label{eq:6}
\end{equation}
where
\begin{equation}
{(\Lambda)^a}_i=\Lambda=
\begin{pmatrix}
1 & 0 & 0 & 0 & 0 & 0 \\
0 & 1 & 0 & 0 & 0 & 0 \\
\end{pmatrix}~,
\end{equation}
and
\begin{equation}
\lambda'_{L}\bar{q}'_{La}\epsilon^{ab}{\Omega_b}^iO'_{Ri}=
\lambda'_{L}\bar{q}'_{La}\epsilon^{ab}{\Omega_b}^i\left({\xi^*_i}^\beta \Sigma_{0\beta\alpha}Q_{R}^\alpha \right),
\label{eq:6bar}
\end{equation}
where
\begin{equation}
{(\Omega)_a}^i=\Omega=
\begin{pmatrix}
1 & 0 & 0 & 0 & 0 & 0 \\
0 & 1 & 0 & 0 & 0 & 0 \\
\end{pmatrix}~.
\end{equation}
The combination of the two interactions breaks the $SU(6)$ global symmetry explicitly. It leads to a potential for the pNGBs at $\mathcal{O}(\lambda_{L}^2\lambda_{L}'^{2})$ of the form
\begin{equation}
\Delta V_s\propto[{(\Lambda)^a}_i{(\Omega^{*})^b}_j\Sigma^{ij}]
[{{(\Omega)_b}^{m}}{(\Lambda^*)_a}^n\Sigma^{*}_{mn}]~,
\end{equation}
which can easily be checked by drawing a one-loop diagram, with $q_L$, $q'_L$, $Q_R$ running in the loop. After expanding it, we obtain a flavon potential
\begin{equation}
\Delta V_s \sim \frac{N_c}{8\pi^2}\lambda_{L}^2\lambda_{L}'^{2} f^4
\left|{(\Lambda)^a}_i{(\Omega^{*})^b}_j\Sigma^{ij}\right|^2
\supset M_s^2|s|^2,
\end{equation}
where
\begin{equation}
M_s^2 \sim \frac{N_c}{8\pi^2}\lambda_{L}^2\lambda_{L}'^2 f^2~.
\end{equation}
Notice that we have chosen different (generations of) elementary quark doublets, $q_L$ and $q_L'$, in the two couplings such that the leading order potential is the quadratic term $|s|^2$.

To have a nontrivial flavon VEV, we want to introduce interactions that explicitly break the $U(1)_F$ symmetry. It can be achieved by mixing $q_L$ to both resonances, which have the quantum number $\mathbf{2}_{1/6,1/2}$ and $\mathbf{2}_{1/6,-1/2}$, with coupling $\lambda_{L}$ and $\lambda_{L}''$. In this way, the loop can be closed at $\mathcal{O}(\lambda_{L}\lambda_{L}'')$ and generate a $s$ tadpole term
\begin{equation}\label{s-tadpole}
\Delta V_s \sim \frac{N_c}{8\pi^2}\lambda_{L}\lambda_{L}'' f^4
\left(\epsilon_{ab}{(\Lambda)^a}_i{(\Omega^{*})^b}_j\Sigma^{ij}\right)
\sim \kappa \,s~,
\end{equation}
where
\begin{equation}
\kappa \sim \frac{N_c}{8\pi^2}\lambda_{L}\lambda_{L}'' g_{\psi}^2 f^3.
\end{equation}

Combining the two potentials we got, the flavon VEV is given by
\begin{equation}\label{VEV}
v_s \sim \frac{\kappa}{M_s^2} \sim \frac{\lambda_{L}\lambda_{L}'' g_{\psi}^2}{\lambda_{L}^2\lambda_{L}'^2} f \propto \lambda_{L}''f,
\end{equation}
which is controlled by the explicit breaking coupling $\lambda_{L}''$. If $\lambda_{L}''$ is small, we can have $v_s\ll f$ with the desired value. Although the tadpole term shifts the vacuum, it preserves the shape of the potential. That is, the masses of the two flavon degrees of freedom, a scalar component $s$ and a pseudoscalar component $a$, are the same with
\begin{equation}
M_s = M_a \sim \sqrt{\frac{N_c}{8\pi^2}}\lambda_{L}\lambda_{L}' f.
\end{equation}
The value is controlled by $\lambda_{L}$ and $\lambda_{L}'$, which can be large and lead to a heavy flavon.

\bibliography{Flavon_Ref}

\end{document}